\documentclass[reprint,showpacs,aps,prl,superscriptaddress]{revtex4-2}
\usepackage{graphicx}
\usepackage{amsmath, amssymb}
\usepackage{mathptmx}
\usepackage{verbatim}
\usepackage{color}
\usepackage{siunitx}
\usepackage{upgreek}
\usepackage{bm}
  \definecolor{YKB}{rgb}{0.00,0.18,0.65}

\begin{document}

\title{Resonant Untrapping of Active Polymers in Breathing Lattices}

\author{Yihang Sun}
\affiliation{School of Physics and Electronics, Hunan University, Changsha 410082, China}
\author{Yixiang Li}
\affiliation{School of Physics and Electronics, Hunan University, Changsha 410082, China}
\author{Tsvi Tlusty}
\affiliation{Department of Physics, Ulsan National Institute of Science and Technology, Ulsan 44919, South Korea}
\author{Guolong Zhu}
\email{zhugl@hnu.edu.cn}
\affiliation{School of Physics and Electronics, Hunan University, Changsha 410082, China}

\date{\today}

\begin{abstract}
In crowded environments, active polymers can trap themselves by winding into long-lived conformations. We show that fluctuations of the surrounding confinement can resonantly accelerate escape from these self-generated traps. Brownian dynamics simulations of a driven semiflexible chain in a breathing obstacle lattice reveal intermittent switching between a compact rotating spiral and an extended translating state. Long-time diffusion increases by up to two orders of magnitude when the environmental fluctuation rate becomes comparable to the spiral’s intrinsic relaxation rate. The enhancement persists under stochastic fluctuations, showing that coherent periodic forcing is not required. Activity creates a second optimum: it promotes escape once favorable conformations form, yet at strong drive stabilizes the spiral and suppresses their formation. Resonant untrapping thus provides a general mechanism by which fluctuating environments regulate transport through barriers generated by internal conformational dynamics.
\end{abstract}

\maketitle

\emph{Introduction---}Transport through crowded environments is often limited by rare escape events from long-lived trapped states \cite{Kurzthaler2021, Irani2022, Mattingly2025, Pietrangeli2025}, which is ubiquitous for active polymers \cite{Schaller2010, Sanchez2011, Sanchez2012, Keber2014, Bianco2018, Locatelli2021, Muzzeddu2024, Das2019, Smrek2017, Deblais2020, Deblais2020_1, Sinaasappel2026, Hooijschuur2026} in living and synthetic soft matter systems. Unlike an externally imposed confining potential \cite{Kramers1940, Hanggi1990, Woillez2019}, these traps arise from activity-induced polymer conformations that are shaped by the surrounding geometry \cite{Mokhtari2019, Chakrabarti2020, Wu2022}. In porous media, previous experimental \cite{Bhattacharjee2019, Heeremans2022, Wei2026, Prathyusha2026, Sinaasappel2025} and theoretical \cite{Sinaasappel2025, Fazelzadeh2023, Zhang2021, Shee2021, Yan2023, Theeyancheri2023, Das2026} studies have established that escape from such trapped states is governed by polymer activity, conformational flexibility, and pore geometry. However, these studies almost exclusively consider static environments, where the escape barrier remains time independent. By contrast, in many biological and biomimetic systems, such as remodeling extracellular matrices \cite{Hynes2009}, contractile tissues \cite{Balasubramaniam2021}, and deformable hydrogel networks \cite{Freeman2018}, the surrounding geometry itself evolves in time, making the escape barrier inherently dynamic. Escape from fluctuating barriers has long been studied for Brownian particles in the framework of resonant activation, which predicts optimal escape when the barrier flipping rate matches the thermal diffusion timescale \cite{Doering1992, Bier1993, Boguna1998}. In these classical systems, the fluctuating barrier is externally imposed and independent of particle dynamics. For active polymers in porous media, however, the trapping barrier emerges from the coupling between activity, polymer conformation, and surrounding geometry, with barrier dynamics governed by the polymer’s internal conformational relaxation. Whether environmental fluctuations can dynamically reshape these emergent barriers and resonantly enhance escape from conformational traps remains unknown.

In this Letter, we address this question by studying the transport of an active polymer in a lattice of pulsating obstacles. Brownian dynamics simulations reveal intermittent polymer dynamics arising from transitions between a compact rotating spiral conformation trapped between neighboring obstacles and an extended linear conformation translating through the gaps. The long-time diffusion coefficient is dominated by the trapping events and depends nonmonotonically on both the obstacle pulsation frequency and the polymer activity strength, varying by up to two orders of magnitude. Theoretical analysis further identifies two conditions for optimal transport. First, the obstacle pulsation rate must become comparable to the relaxation rate of the polymer, enabling resonant escape from the trapped state. The resonance therefore couples an environmental timescale to an intrinsic conformational mode of the polymer. Indeed, the same optimum persists when periodic breathing is replaced by stochastic environmental fluctuations. Second, the polymer activity strength must balance the competition between the formation of configurations favorable for escape and its ability to lower the escape barrier once such configurations are formed. Activity thus plays a dual role: it stabilizes the trapped conformation while also driving escape from it. These results establish a general mechanism for controlling the transport of biological active polymers in fluctuating environments.

\emph{Simulation.---}In this work, Brownian dynamics simulations are employed to investigate the diffusion of an active polymer in a two-dimensional lattice of pulsating obstacles \footnote{See Supplemental Material for details of simulation model and methods, derivation of hopping time and hopping length, nonequilibrium escape barrier, resonant untrapping under spatial and temporal disorder, and supplementary movies, which includes Refs.~\cite{Nikoubashman2017, Best2010, Best2005, Hummer2004, Manna2017}.}. \nocite{Nikoubashman2017, Best2010, Best2005, Hummer2004, Manna2017} The polymer is modeled as a semiflexible bead-spring chain composed of $N$ monomers, each with diameter $b$. Chain connectivity is enforced by a finitely extensible nonlinear elastic potential, while bending rigidity is incorporated via a harmonic angle potential. Excluded-volume interactions between monomers are modeled by the Weeks–Chandler–Andersen potential. The polymer moves in an ordered porous medium formed by fixed obstacles arranged on a square lattice [Fig.~1(a)]. The obstacle diameter pulsates in time as
\begin{equation}
    d_\mathrm{obs}(t) = d_0 + d_1 \cdot \sin(2\pi f t).
\end{equation}
Polymer--obstacle overlap is prevented by a truncated harmonic repulsion. Each monomer, except for the head and tail monomers, is driven by an active force directed along the local tangent with fixed magnitude $F_{ \mathrm{act}}$. The dynamics of the monomers is described by the overdamped Langevin equation:
\begin{equation}
    \gamma \dot{\mathbf{r}}_i = F_{ \mathrm{act}} \cdot \frac{\mathbf{r}_{i+1}-\mathbf{r}_{i-1}}{\left|\mathbf{r}_{i+1}-\mathbf{r}_{i-1}\right|} - \nabla_i U + \boldsymbol{\xi}_i,
\end{equation}
where $\gamma$ is the friction coefficient, $U$ is the total potential energy and $\boldsymbol{\xi}$ is the thermal noise. We measure length in units of $b$, energy in units of $k_\mathrm{B} T$, and time in units of $\tau_0=b^2 / D_0$, where $D_0=k_\mathrm{B} T / \gamma$ is the thermal diffusion coefficient of an isolated monomer. The diffusion of the active polymer is investigated by varying two key control parameters: the polymer activity strength, quantified by the P\'{e}clet number $\text{Pe} \equiv F_{\mathrm{act}} b / k_\mathrm{B} T$, and the obstacle pulsation frequency, nondimensionalized as $\emph{f}\tau_0$. The resonant untrapping phenomenon reported below remains robust over a broad parameter range, including variations in chain length $N$, persistence length $l_\mathrm{p}$, obstacle diameter $d_0$, pulsation amplitude $d_1$, and obstacle area fraction $\phi$ [Fig.~S1]. Unless otherwise specified, we set $N=30$, $l_\mathrm{p}/Nb=0.5$, $d_0=10b$, $d_1=0.5b$, and $\phi \equiv {\pi d_0^2}/{(4L^2)}=0.6$, where $L$ is the lattice spacing.

\begin{figure}[t]
\centering
\includegraphics[width=8.6cm]{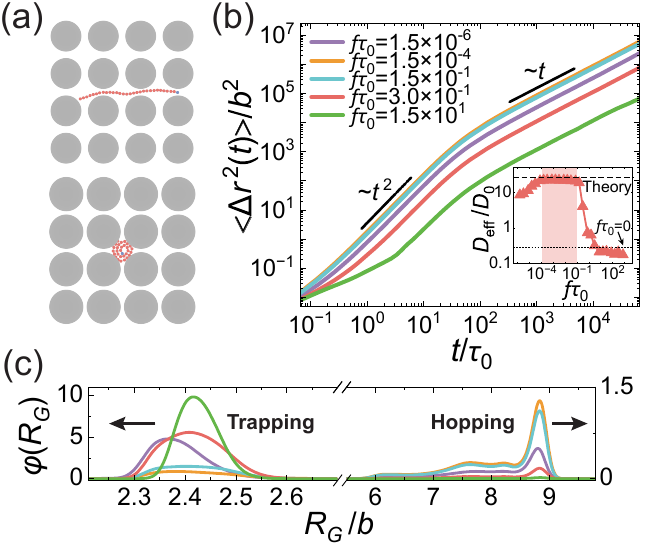}
\caption{(a) Schematics of an active polymer in a lattice of pulsating obstacles, with the blue bead denoting the polymer head. (b) Center-of-mass mean-square displacement $\langle \Delta r^2(t) \rangle$ of the polymer for different pulsation frequencies $f$. Inset: normalized diffusion coefficient $D_\mathrm{eff}$ as a function of $f$, where the dashed and dotted lines denote the theoretical upper bound of $D_\mathrm{theory}$ [Eq.~(7)] and $D_\mathrm{eff}$ at $f=0$, respectively. (c) Probability distribution of the radius of gyration $R_G$ for different $f$.}
\end{figure}

We begin by examining the effect of the pulsation frequency $f$ on polymer diffusion by computing the center-of-mass mean-square displacement $\langle\Delta r^2(t)\rangle$ at fixed $\text {Pe}=1.8$ [Fig.~1(b)]. Consistent with the behavior of an active Brownian particle \cite{Bechinger2016}, the polymer motion is diffusive at short time scales, ballistic at intermediate time scales, and again diffusive at long time scales. Interestingly, the long-time diffusion coefficient $D_\mathrm{eff}$ depends nonmonotonically on $f$ [inset of Fig.~1(b)]. For $\emph{f}\tau_0 \gtrsim 10^{-1}$, decreasing $f$ leads to a substantial increase in the diffusion coefficient, rising by up to two orders of magnitude. For $10^{-4} \lesssim \emph{f}\tau_0 \lesssim 10^{-1}$, the diffusion coefficient remains nearly constant, forming a high-value plateau. For $\emph{f}\tau_0 \lesssim 10^{-4}$, further decreasing $f$ reduces the diffusion coefficient. Compared with the static case ($f=0$), diffusion is slightly suppressed at high $f$ but significantly enhanced at low $f$. To uncover the underlying mechanism of the diffusion behavior, we compute the radius of gyration $R_G$ of the polymer. As shown in Fig.~1(c), the probability distribution of $R_G$ exhibits two prominent peaks, corresponding to a compact spiral conformation when the polymer is trapped in a cavity formed by four neighboring obstacles, and an extended linear conformation when it passes through the channels connecting adjacent cavities, respectively [Fig.~1(a)]. At intermediate $f$, the increased probability of the extended conformation indicates that the polymer spends more time moving between obstacles, thereby enhancing the long-time diffusion.

\begin{figure}
\centering
\includegraphics[width=8.6cm]{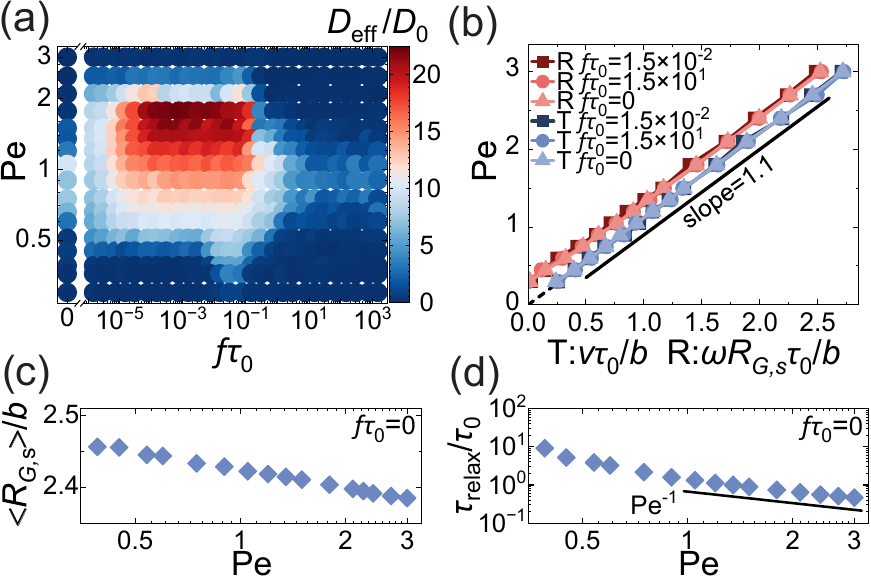}
\caption{(a) Phase diagram of $D_\mathrm{eff}$ in the $\text{Pe}-f$ plane. (b) $\mathrm{Pe}$ versus the translational (T) velocity of the linear conformation and the rotational (R) velocity of the spiral conformation for different $f$. (c) The gyration radius of the spiral conformation $R_{G,\mathrm{s}}$ as a function of $\mathrm{Pe}$. (d) Relaxation time of the spiral conformation as a function of $\mathrm{Pe}$.}
\end{figure}

Subsequently, we evaluate $\langle \Delta r^2(t) \rangle$ over a range of $\mathrm{Pe}$ and construct the phase diagram of $D_\mathrm{eff}$ as a function of $\mathrm{Pe}$ and $f$ [Fig.~2(a)]. The qualitative dependence of $D_\mathrm{eff}$ on $f$ remains largely unchanged, particularly at intermediate $\mathrm{Pe}$ [Fig.~S2(a)]. However, a counterintuitive nonmonotonic dependence of $D_\mathrm{eff}$ on $\mathrm{Pe}$ is observed, which is robust against variations in $f$ [Fig.~S2(b)]. As $\mathrm{Pe}$ increases, $D_\mathrm{eff}$ initially increases due to the activity-enhanced translational motion of the polymer in the linear conformation [Fig.~2(b)]. Strikingly, further increasing $\mathrm{Pe}$ leads to a dramatic reduction in $D_\mathrm{eff}$, indicating a dominant role of the spiral conformation at large $\mathrm{Pe}$. We therefore examine the effect of $\mathrm{Pe}$ on the structural and dynamic characteristics of the spiral conformation. As shown in Fig.~2(c), the gyration radius of the spiral conformation $R_{G,\mathrm{s}}$ decreases monotonically with increasing $\mathrm{Pe}$, indicating a more compact spiral conformation. The characteristic relaxation time of the spiral, defined as the time when the autocorrelation function $C(t) = \langle \delta R_{G,\mathrm{s}}(t_0+t)\,\delta R_{G,\mathrm{s}}(t_0)\rangle / \langle \delta R_{G,\mathrm{s}}^2\rangle$ decays to zero, scales as $\tau_\mathrm{relax} \sim \mathrm{Pe}^{-1}$ [Fig.~2(d)]. Furthermore, we compute the rotational velocity of the spiral conformation and observe a linear increase with $\mathrm{Pe}$ [Fig.~2(b)]. However, the linear relationship exhibits a nonzero intercept, suggesting an additional dissipation pathway beyond rotational motion. Specifically, a fraction of injected active energy is consumed in stabilizing the spiral conformation.

\begin{figure}[t]
\centering
\includegraphics[width=8.6cm]{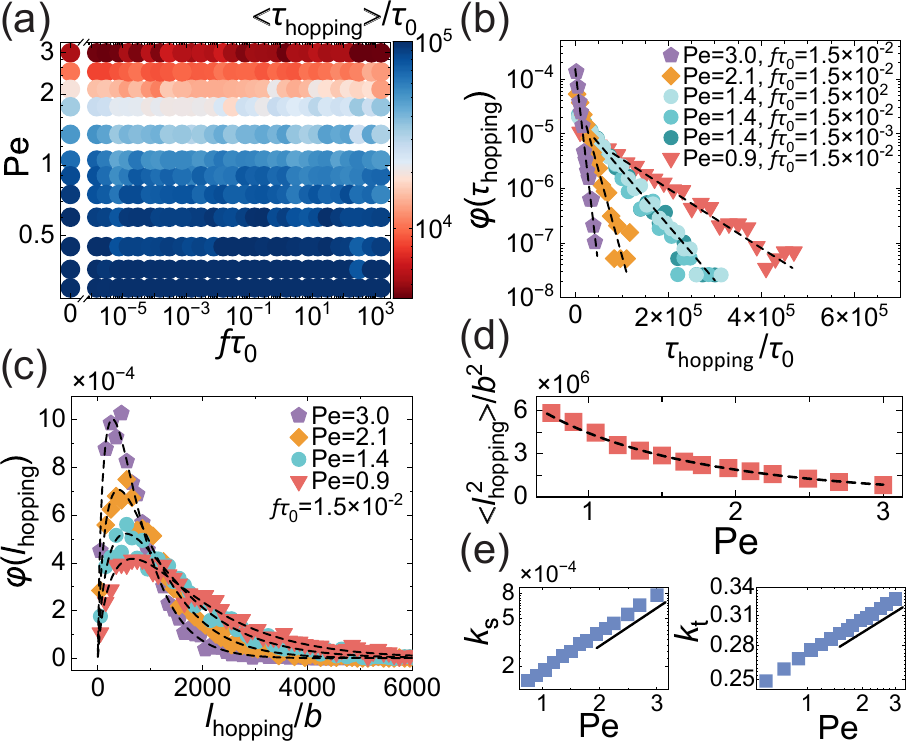}
\caption{(a) Phase diagram of mean hopping time in the $\text{Pe}-f$ plane. (b-c) Distributions of hopping time (b) and hopping length (c) for different $\mathrm{Pe}$ and $f$. (d) Mean square hopping length as a function of $\mathrm{Pe}$. Dashed lines in (b), (c), and (d) represent theoretical results from Eqs.~(3), (4), and (5), respectively. (e) Spiraling probability $k_\mathrm{s}$ and reorientation probability $k_\mathrm{t}$ as functions of $\mathrm{Pe}$.}
\end{figure}

\emph{Theoretical analysis---}As mentioned above and shown in the simulation movies \footnotemark[\value{footnote}], the polymer switches between a translationally immobilized, rotating spiral conformation and a fast-moving linear conformation, resulting in intermittent dynamics consisting of trapping and hopping events. To uncover the physical origin of the unexpected dependence of $D_{\mathrm{eff}}$ on $\mathrm{Pe}$ and $f$, we therefore analyze the motion within the continuous-time random-walk (CTRW) framework \cite{Montroll1965}, where the trapping-time distribution and hopping-length statistics provide the central dynamical descriptors.

First, we extract the hopping times $\tau_\mathrm{hopping}$ from the trajectories and find that $\langle \tau_\mathrm{hopping} \rangle$ is essentially independent of $f$ but decreases with increasing $\mathrm{Pe}$ [Fig.~3(a)]. After visiting a cavity, the polymer may enter the spiral state and terminate the current hopping event. The spiraling probability is assumed to be a uniform constant $k_\mathrm{s}$ because all cavities provide equivalent environments. The hopping time distribution is therefore derived as \footnotemark[\value{footnote}]
\begin{equation}
\upvarphi(\tau_\mathrm{hopping}) = \frac{ k_\mathrm{s} F_{\mathrm{act}} }{L \gamma_{\mathrm{eff}}} \exp \left( -\frac{ k_\mathrm{s} F_{\mathrm{act}} }{L \gamma_{\mathrm{eff}}} \cdot \tau_\mathrm{hopping} \right).
\end{equation}
Here, $\gamma_{\mathrm{eff}} \approx 1.1 \gamma$ is the effective friction coefficient, arising from intrachain coupling and conformational constraints [Fig.~2(b)]. The exponential form is clearly confirmed by the data shown in Fig.~3(b). When passing each cavity, the polymer either proceeds forward or reorients perpendicular to its incoming direction, while the vanishingly rare backward reorientation is neglected. Accordingly, the reorientation probability in each cavity is denoted by $k_\mathrm{t}$, with equal probabilities for leftward and rightward reorientations. Consequently, the hopping diffusion coefficient is given by $D_\mathrm{hopping}= (2-k_\mathrm{t}) F_\mathrm{act} L / 4 k_\mathrm{t} \gamma_\mathrm{eff}$ \footnotemark[\value{footnote}]. The theoretical distribution of hopping length, supported by the data presented in Fig.~3(c), is therefore derived as
\begin{equation}
\upvarphi(l_\mathrm{hopping})=\frac{ l_\mathrm{hopping} 4k_{\mathrm{s}} k_\mathrm{t} }{ (2-k_\mathrm{t}) L^2 } K_0\left(l_\mathrm{hopping} \sqrt{ \frac{ 4k_{\mathrm{s}} k_\mathrm{t} }{ (2-k_\mathrm{t}) L^2 } }\right),
\end{equation}
where $K_0$ denotes the zeroth-order modified Bessel function of the second kind. Furthermore, we can derive the mean-square hopping length as [Fig.~3(d)]
\begin{equation}
\langle l_\mathrm{hopping}^2 \rangle = \frac{ (2-k_\mathrm{t}) L^2 }{ k_{\mathrm{s}} k_\mathrm{t} }.
\end{equation}
This expression shows that activity controls the hopping length by modulating $k_\mathrm{s}$ and $k_\mathrm{t}$, which are obtained directly from the trajectories. Specifically, increasing activity enhances conformational flexibility, facilitating spiraling and reorientation. Within the range of $\mathrm{Pe}$ examined, $k_\mathrm{s}$ exhibits an exponential dependence on $\mathrm{Pe}$, whereas $k_\mathrm{t}$ follows a power-law scaling with $\mathrm{Pe}$ [Fig.~3(e)]. The good agreement between the theoretical description and the simulation results in Figs.~3(b)-3(d) corroborates again the validity of this theoretical model.

Next, we turn to the trapping times $\tau_\mathrm{trapping}$ in the spiral state.  As shown in Fig.~4(a), $\langle\tau_\mathrm{trapping}\rangle$ exhibits essentially the same dependence on $\mathrm{Pe}$ and $f$ as $D_{\mathrm{eff}}$ [Fig.~2(a)], indicating that trapping events play the dominant role in controlling polymer transport. Moreover, $\tau_\mathrm{trapping}$ follows an exponential distribution [Fig.~4(b)], which yields a finite $\langle\tau_\mathrm{trapping}\rangle$, and together with the finite $\langle \tau_\mathrm{hopping} \rangle$ and $\langle l_\mathrm{hopping}^2 \rangle$, implies normal diffusion at long times within the CTRW framework \cite{Montroll1965}, consistent with our numerical observations. Before escaping a trap, the polymer undergoes an unspiraling process, which requires the monomers near the head to adopt a curvature opposite to that of the spiral [Fig.~4(c)]. This step involves surmounting an associated energy barrier. To quantify the underlying equilibrium barrier, a biased-sampling approach \cite{Frenkel2023, Zwanzig1954, Liu1996} is implemented to obtain the free energy along the reaction coordinate $\chi$, which is defined as the integrated curvature of the chain. This reaction coordinate is proven to be a robust descriptor of the relevant free energy landscape [Fig.~S3(a)]. In Fig.~4(d), the reaction coordinate is shifted so that $\chi=0$ corresponds to the critical configuration at which the local curvature near the polymer head changes sign [Fig.~4(c)].

In the presence of activity, the escape barrier is controlled by two main contributions. On the one hand, the stable spiral conformation, of which the compactness dominates the equilibrium energy barrier, is controlled by $\mathrm{Pe}$. Specifically, increasing $\mathrm{Pe}$ leads to a more tightly wound spiral [Fig.~2(c)], corresponding to a larger entropic trapping barrier. Moreover, the compactness of the spiral depends sensitively on the pulsation phase. When the obstacle radius is maximal, which forces the polymer to adopt the most compact spiral conformation, the barrier $V_{\mathrm{eq}}^+$ is maximal and shows no dependence on $\mathrm{Pe}$ [Fig.~4(e)]. In contrast, when the obstacle radius is minimal, which produces the most extended spiral conformation, the barrier $V_{\mathrm{eq}}^-$ is minimal, and increasing $\mathrm{Pe}$ raises the barrier [Fig.~4(e)]. On the other hand, activity might modify the equilibrium escape barrier through active energy input, which is detailed in Sec.~III in \footnotemark[\value{footnote}]. Briefly, this modification can be separated into two regimes along the reaction coordinate [Fig.~4(c-d)]. When $\chi<0$, escape-favorable configurations have not yet formed, and the tangential active force stabilizes the spiral and suppresses their formation, thereby raising the equilibrium energy barrier $V_{\mathrm{eq,1}}$ \cite{Woillez2019}. Once these configurations are formed, corresponding to $\chi>0$, the active force is aligned with the escape pathway and lowers the remaining barrier $V_{\mathrm{eq,2}}$.

\begin{figure}
\centering
\includegraphics[width=8.6cm]{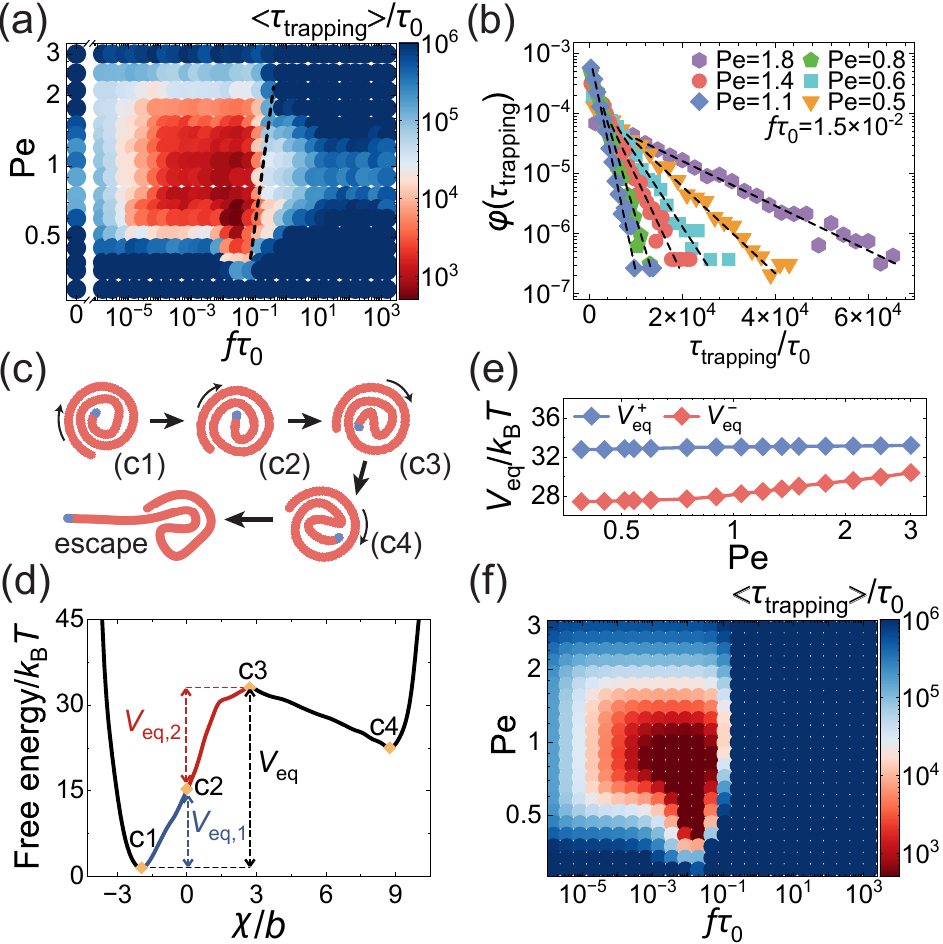}
\caption{(a) Phase diagram of mean trapping time in the $\text{Pe}-f$ plane. The dashed line represents $2 \pi f \tau_\mathrm{relax}=1$. (b) Distribution of trapping time for different $\mathrm{Pe}$. (c) Snapshots illustrating the escape process. Curved arrows indicate the rotation direction. (d) Equilibrium free energy as a function of the reaction coordinate $\chi$. (e) The maximal and minimal equilibrium energy barriers as a function of $\mathrm{Pe}$. (f) Theoretical phase diagram of the mean escape time.}
\end{figure}

Finally, the modified nonequilibrium escape barriers $V_{\mathrm{neq}}^{\pm}$ are modulated by obstacle pulsations. Since conformational rearrangements of the polymer require a finite relaxation time, the instantaneous barrier may not fully reach $V_{\mathrm{neq}}$ within a pulsation cycle. As a result, the effective barrier is governed by the interplay between the pulsation frequency and the intrinsic relaxation time of the spiral conformation. Following the Maxwell relaxation model \cite{Rubinstein2003}, the difference between the two effective energy barriers is reduced to $V_{\mathrm{eff}}^+ - V_{\mathrm{eff}}^- =\left(V_{\mathrm{neq}}^+ - V_{\mathrm{neq}}^-\right) / \sqrt{1+\left(2\pi f \tau_{\mathrm{relax}}\right)^2}$. In particular, $V_{\mathrm{eff}}^+ = V_{\mathrm{neq}}^+$ because the polymer is forced to adopt a compact spiral conformation when the obstacle radius reaches its maximum. Thus far, we have provided the effective energy barriers associated with the escape process. However, the time-dependent barrier prevents a direct application of the standard Kramers escape picture. We therefore adopt a two-state approximation considering activated escape only through the highest and lowest energy barriers. The probability densities of the active polymer along the reaction coordinate $\rho^\pm(\chi)$ satisfy the coupled Fokker-Planck equations:
\begin{equation}
\frac{\partial}{\partial t}\binom{\rho^{+}}{\rho^{-}}=\left(\begin{array}{cc}
-2f + \mathcal{L}^{+} & 2f \\
2f & -2f+ \mathcal{L}^{-}
\end{array}\right)\binom{\rho^{+}}{\rho^{-}}.
\end{equation}
Here $\mathcal{L}^{\pm}$ is the Fokker-Planck operator, defined as $\frac{\partial}{\partial \chi}\left\{ \frac{ D_{\chi} }{ k_\mathrm{B} T } \frac{\partial V_{\text{eff}}^{\pm}}{\partial \chi} + D_{\chi} \frac{\partial}{\partial \chi}\right\}$, and $D_{\chi}=8.5D_0$ is the diffusion coefficient of the reaction coordinate, obtained by its short-time variance [Fig.~S3(b)]. The mean escape time predicted by $V_{\text{eff}}^{\pm}$ is shown in Fig.~4(f), which quantitatively reproduces the phase diagram of $\tau_\mathrm{trapping}$.

Within the CTRW framework, the theoretical diffusion coefficient is given by \cite{Montroll1965}
\begin{equation}
D_\mathrm{theory} = D_\mathrm{hopping} \cdot \frac{\langle\tau_\mathrm{hopping}\rangle}{\langle\tau_\mathrm{trapping}\rangle + \langle\tau_\mathrm{hopping}\rangle}.
\end{equation}
When $\langle\tau_\mathrm{trapping}\rangle \ll \langle\tau_\mathrm{hopping}\rangle$, the theory predicts an upper bound of $D_\mathrm{theory} = D_\mathrm{hopping}$, which is in quantitative agreement with the numerical result in the inset of Fig.~1(b) and in Fig.~S2(a). To reach the bound, the trapping time should be as short as possible. However, the dependence of $\langle\tau_\mathrm{trapping}\rangle$ on $\mathrm{Pe}$ and $f$ is intricate. At large $\mathrm{Pe}$, the formation of escape-favorable configurations is suppressed by activity, whereas at small $\mathrm{Pe}$, the barrier-lowering effect is limited. Thus activity plays a dual role: it stabilizes the conformational trap while also driving escape from it. The competition between these two mechanisms results in the observed nonmonotonic dependence of $\langle \tau_\mathrm{trapping} \rangle$ and $D_{\mathrm{eff}}$ on $\mathrm{Pe}$, although $D_\mathrm{hopping}$ increases monotonically with $\mathrm{Pe}$ [Fig.~S2(a)]. At high $f$, the polymer cannot relax sufficiently, and escapes only through the higher energy barrier. At intermediate $f$, resonant activation arises from the matching between the pulsation of obstacles and the relaxation of the polymer, resulting in fast escape \cite{Doering1992, Bier1993, Boguna1998}. The predicted onset of resonance at $2 \pi f = \tau_\mathrm{relax}^{-1} \sim \mathrm{Pe}$ [Fig.~2(e)] is consistent with the numerical results, highlighted by the dashed line in Fig.~4(a). We further demonstrate this timescale matching by tuning $\tau_{\rm relax}$ through variations in $l_\mathrm{p}$, showing that the onset frequency of resonant untrapping consistently satisfies $2 \pi f \tau_\mathrm{relax}=1$ [Fig.~S1(b)]. At low $f$, the escape becomes sensitive to the phase of the pulsation and the polymer must wait for a sufficiently low escape barrier, leading to the increase of $\langle\tau_\mathrm{trapping}\rangle$.

\emph{Stochastic fluctuation---}The synchronized breathing of ordered obstacles considered above represents an idealized form of environmental fluctuations. To examine the generality of the resonant untrapping mechanism, we introduce spatial and temporal disorder by considering random obstacle positions and random pulsation phases, both of which preserve the characteristic nonmonotonic dependence of $D_\mathrm{eff}$ on $f$ [Fig.~S4]. Furthermore, we replace the deterministic periodic breathing by a stochastic process with a finite correlation time. Specifically, the obstacle diameter is given by [Fig.~5(a)]
\begin{equation}
    d_\mathrm{obs}(t) = d_0 + d_1 \cdot \tanh[X(t)],
\end{equation}
where $X(t)$ obeys an Ornstein--Uhlenbeck process,
\begin{equation}
    \dot X = - X / \tau_\mathrm{OU} + \sqrt{ 2 \sigma_\mathrm{R}^2 / \tau_\mathrm{OU} }  \xi(t).
\end{equation}
Here, $\tau_\mathrm{OU}$ is the correlation time and $\xi$ is the thermal noise. The diffusion coefficient again exhibits a pronounced maximum at an intermediate correlation time [Fig.~5(b)], demonstrating that resonant untrapping does not rely on coherent periodic driving but is instead governed by the interplay between the environmental fluctuation timescale and the intrinsic relaxation dynamics of the polymer.

\begin{figure}
\includegraphics[width=8.6cm]{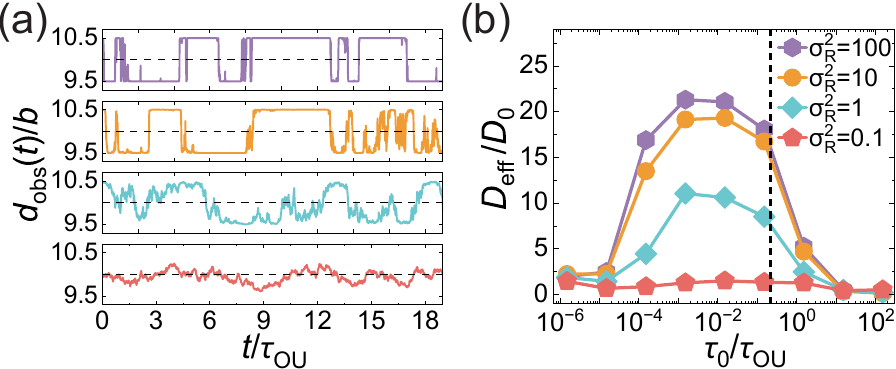}
\caption{(a) Obstacle diameter as a function of time for different noise variance $\sigma_\mathrm{R}^2$. (b) $D_\mathrm{eff}$ as a function of $\tau_\mathrm{OU}$. The dashed line indicates $2 \pi \tau_\mathrm{OU}^{-1} \tau_\mathrm{relax}=1$.}
\end{figure}

\emph{Conclusion---}We have demonstrated resonant untrapping of active polymers in a lattice of pulsating obstacles, accelerating their transport by up to two orders of magnitude. Combining Brownian dynamics simulations and theoretical analysis, we elucidate the mechanisms underlying the nonmonotonic dependence of polymer diffusion on obstacle pulsation frequency and polymer activity strength. Optimal transport occurs when environmental fluctuations match the intrinsic relaxation dynamics of the polymer and when activity balances the formation of escape-favorable configurations and the barrier-lowering effect. More broadly, fluctuating environments can regulate active transport by coupling to barriers that emerge from internal conformational dynamics.

\emph{Acknowledgments---}We acknowledge support from the National Natural Science Foundation of China (Grant No. 22303029) and the Natural Science Foundation of Hunan Province (Grant No. 2025JJ60015). T.T. acknowledges support from the National Research Foundation of Korea under Grant No. NRF-RS-2025-00573354, the InnoCORE Bio-MAX program, and the U.S. Office of Naval Research under Award No. N00014-26-1-2401.

\emph{Data availability---}The data that support the findings of this article are openly available.


\bibliography{Bibliography}

\end{document}